\documentclass[preprint,aps,prb,showpacs,preprintnumbers,amsmath,amssymb,superscriptaddress]{revtex4}

\usepackage{graphicx}% Include figure files

\usepackage{natbib}

\begin{document}
\title{Universal Limits of Thermopower and Figure of Merit from Transport Energy Statistics}

\newcommand\affHZDR{\affiliation{%
Helmholtz-Zentrum Dresden - Rossendorf,
Institute of Ion Beam Physics and Materials Research,
Bautzner Landstra{\ss}e 400, 01328 Dresden, Germany}
}
\newcommand\affTUC{\affiliation{%
Institute of Physics, TU Chemnitz,
09107 Chemnitz, Germany}
}
\newcommand\affNanoNet{\affiliation{%
Helmholtz-Zentrum Dresden - Rossendorf,
International Helmholtz Research School for Nanoelectronic Networks (IHRS NanoNet),
Bautzner Landstra{\ss}e 400, 01328 Dresden, Germany}
}
\newcommand\affDCMS{\affiliation{
Dresden Center for Computational Materials Science (DCMS),
TU Dresden, 01062 Dresden, Germany}
}
\newcommand\affCFAED{\affiliation{
Center for Advancing Electronics Dresden (cfaed),
TU Dresden, 01062 Dresden, Germany}
}
\author{Peter~Zahn}
\email{p.zahn@hzdr.de}
\affHZDR\affNanoNet\affDCMS
%\author{Sibylle~Gemming}
%\email{s.gemming@hzdr.de}
%\affHZDR\affTUC\affNanoNet\affDCMS\affCFAED

\date{\today}% It is always \today, but any date may be explicitly specified

\begin{abstract}
The search for new thermoelectric materials aims at improving their
power and efficiency, as expressed by thermopower $S$ and figure of
merit $ZT$. 
By considering a very general transport spectral function $W(E)$,
expressions for $S$ and $ZT$ can be derived, which contain the
statistical weights of an effective distribution function only, see
Refs. \cite{eltschka16,mahan96,matveev00}. 

We assumption of a Lorentzian shape with width $k_BT$
resulting from the electron-phonon coupling allows to estimate an
upper limit of $S$ and $ZT$ independent on the microscopic mechanisms of the
transport process.
A simple estimate for an upper limit of the thermopwer $S$ is derived
from {} formula. It is given by 3 times the unit of the thermopower
$k_b/e$ which is about 250~$\mu V/K$.

We consider different systems which represent the general features 
of the electronic structure of thermoelectric relevant materials very
well. 
The transport integrals were evaluated varying the band gap size
and the chemical potential position. 
For all cases upper limits for
both, the thermopower and the figure of merit, are obtained. 
The universal limit of $|S|$ is given by 1.88 in units of
$k_B/e$, which is about 160~$\mu V/K$.
The universal limit for $ZT$ is obtained by about 1.11, which is in good
agreement with available thermoelectric systems and devices.

\end{abstract}

%\pacs{}% PACS, the Physics and Astronomy Classification Scheme.
%\keywords{Suggested keywords}% For display use 'showkeys' class option

\maketitle

 \section{Introduction}
  % Introduction
%
%
Limitations of TE applications and efficiency is given by the Figure
of Merit $ZT$. 
Ref. \cite{vining09} pointed to the limit range of obtained ZT values,
but no reasons were given for the pessimistic view that a ZT of 2
is 'eventually plausible', and a ZT of 4 'is ambitious'.
Some reports on ZT values larger than 2 are available, e.g.
3.5 in quantum dot superlattices \cite{harman05}, 
2.4 and 2.9 in bismuth telluride and antimonide superlattices
\cite{rama01}, 
and 2.2 in antimony silver telluride materials \cite{hsu04}.

Here we give a simple explanation for the limitations of achievable ZT
values.
We assume the presence of electron-phonon coupling which broadens the
width of all electronic states by $k_BT$. This broadening along the
real energy axis is besides the limits in the electron life time which
gives a shift of the electron density resonances along the imaginary
energy axis. 

To study the transport properties we consider the transport spectral
function as a superposition of Lorentzian peaks.
So, we assume that all features of the transport spectral function are
broadened by $k_BT$.

 \section{Methods and Assumptions}
     \subsection{Thermoelectric Transport Coefficients}

Following Refs. \cite{mahan96,hinsche16,eltschka16} we write the
linear transport formalism using a general 
{\bf Transport spectral function $w(E)$}. 
This is not restricted to diffusive transport and 
solving a quasi-classical Boltzmann equation.
A similar approach was discussed by \cite{matveev00} generalizing it
to multi-terminal geometries.
The transport spectral function $w(E)$ quantifies the
contribution of the electronic states at energy $E$
to the electrical current if different 
occupations in the terminals drive a current through the device.

Longitudinal transport with 
electric field ${\cal E}$, 
temperature gradient $\nabla T$, and 
electrical and thermal currents $j$ and $Q$, respectively, 
in the same direction will be considered in the following:
%
%\begin{eqnarray*}
\begin{align}
j &= e^2 I_0 {\cal E}  - \frac{e}{T} I_1 \nabla T
=
\sigma {\cal E} -\sigma S \nabla T 
\\
Q &= e I_1 {\cal E} - \frac{1}{T} I_2 \nabla T
=
\Pi j -\kappa \nabla T 
\quad .
\end{align}
%\end{eqnarray*}
%
The integrals $I_n$ over the transport spectral function
$w(E)$ depend explicitely on the chemical potential $\mu$ and the
temperature $T$, and are given by
\begin{align}
I_n(\mu, T) 
&=
\int_{-\infty}^{+\infty} dE
\; W(E) \;
\left[ - \frac{\partial f_{FD}(\mu,T)}{\partial E}\right]
 (E-\mu)^n
\\
I_n &=
(k_B T)^n
\int_{-\infty}^{+\infty} dE
\; W(\epsilon) \;
\frac{e^\epsilon}{\left( 1+e^\epsilon \right)^2}
 \epsilon^n
\quad \text{with} \quad
\epsilon=\frac{E-\mu}{k_B T}
\quad .
\end{align}
Using the conductivity $\sigma=e^2 I_0$, 
the thermopower $S=-\frac{1}{|e|T} \frac{I_1}{I_0}=
-\frac{k_B}{|e|} \frac{<\epsilon>}{k_BT}$,
the Peltier coefficient $\Pi = S T = - \frac{<\epsilon>}{|e|}$,
and the electron thermal conductivity 
$\kappa / \sigma = \frac{1}{T\sigma} - T S^2 =
\frac{1}{e^2 T} \left( <\epsilon^2> - 
<\epsilon>^2 \right)$, 
we arrive at the main result:
\begin{equation}
{\boldmath
ZT = 
\frac{\sigma S^2}{\kappa} T
=
\frac{<\epsilon>^2}
{<\epsilon^2> - <\epsilon>^2}
}
\quad .
\end{equation}
Here we neglected the contribution of the lattice thermal conductivity
$\kappa_L$, so we derive an upper limit for 
the figure of merit $ZT$.
To simplify the discussion and the calculations in the following, we
consider all energies in units of $k_B T$. So, $\epsilon$ is given in
units of 1 and measures the energy relative to the chemical
potential. The {\sc Fermi}-{\sc Dirac} occupation function reads like 
$1/\left( 1+\exp(x)\right)$ in these units.
  \subsection{Electron-Phonon Coupling}
Now, the general properties of the transport spectral function should
be introduced as the {\bf Main Assumption}:
\\
All features of the transport spectral function $W(E)$
are {\bf broadened by electron-phonon coupling}
with a Lorentzian line shape.
This results in a superposition of Lorentzian peaks
of width ${\boldmath \Gamma=k_B T}$.
The most important consequence with respect to the thermoelectric
performance is the smooth behavior of the transport spectral function
on the scale of the thermal energy $k_BT$. 
As we will show, this limits the thermopower and the figure of merit to
certain values independent on the transport mechanism and the
dimensionality of the system.
  \subsection{Systems under Consideration}
\begin{figure}[hbt]
\centering
\includegraphics[width=0.9\columnwidth]{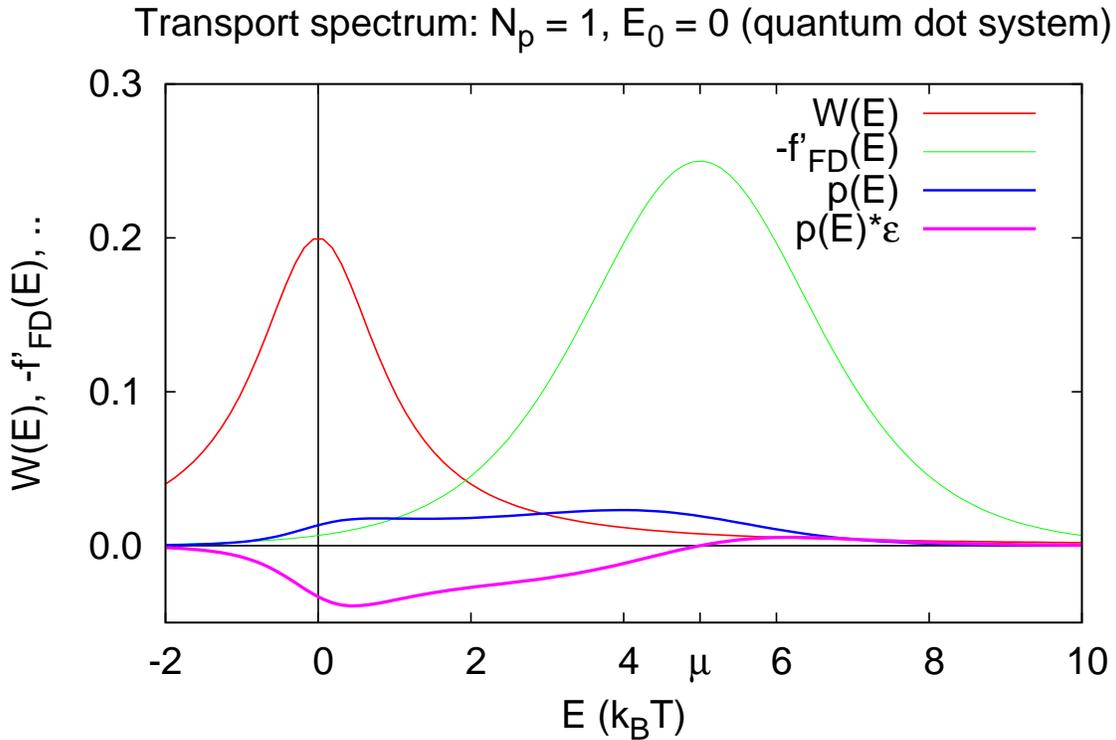}
\caption{\label{fig:s1} 
Quantum dot system: Transport spectral function $W(E)$, derivative of Fermi
occupation function $-f'_{FD}$ for $\mu=4.8$, the effective
distribution function $p(\epsilon)$, and the product
$p(\epsilon)\times\epsilon$ to visualize the contributions to
$<\epsilon>$. The energy $E$ and the relative energy $\epsilon=E-\mu$
are given in units of the thermal energy $k_BT$.
}
\end{figure}

Assuming the above mentioned level broadening all transport spectral
functions can be formed by a superposition of Lorentzian peaks at
different energetic positions and with different weights. 
\\
1/ The first system to consider is a single level system. It might be a
quantum dot with one energy level $E_0$ coupled to two electrodes
\begin{equation}
W(E) = \frac {1}{1+(E-E_0)^2}
\quad.
\end{equation}
All energies are given in units of $k_BT$.
This is illustrated in fig.~\ref{fig:s1}.

2/ The next system under consideration contains 2 levels with spacing 
$E_1-E_0$ to 
mimic a band gap material. 
For simplicity the peak position $E_0=0$ will define the zero of the
energy scale.
Both peaks have equal weight $W_0=W_1=1$:
\begin{equation}
W(E) = \frac {1}{1+(E)^2} + \frac {1}{1+(E-E_1)^2}
\quad.
\end{equation}

3/ To mimic more sophisticated spectral functions the 2 level system
with different peak weights $W_0$ and $W_1$ will be considered. For
simplicity peak width $W_0=1$ is set to one.
\begin{equation}
W(E) = \frac {1}{1+(E)^2} + \frac {W_1}{1+(E-E_1)^2}
\quad.
\end{equation}
By considering different separations $E_1$ and peak weight ratios $W_1$
a wide range of functional behaviors can be simulated. As under the
defined circumstances every transport spectral function is a linear
superposition of Lorentzian peaks, this investigate clearly shows that 
by superposition no larger values for thermopower and figure of merit
can be obtained.

\begin{figure}[hbt]
\centering
\includegraphics[width=0.9\columnwidth]{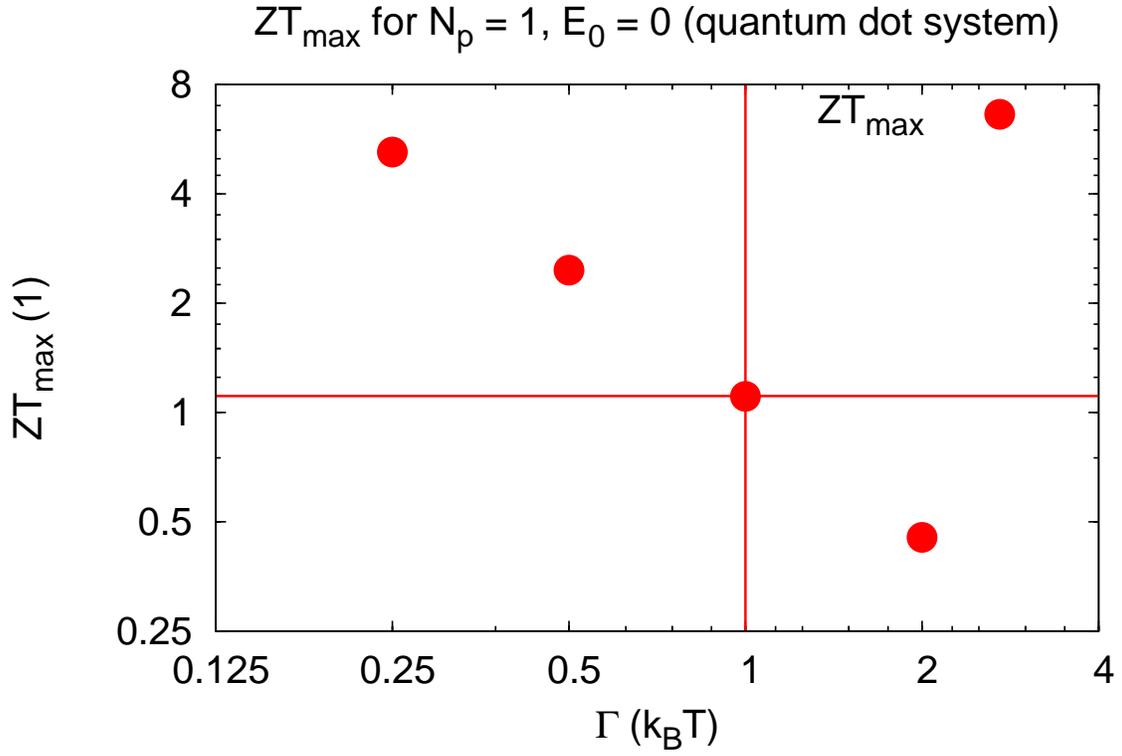}
\caption{\label{fig:ZT-max-GG}
Quantum dot system: 
Maximum $ZT$ as function of coupling strength $\Gamma$.
For small $\Gamma$ the increase of $ZT$ is roughly $\Gamma^{-5/4}$.
}
\end{figure}

4/ This case studies the behavior for a reduced peak width $\Gamma=N
\times k_B T$ with $N<1$. It is not considered by which means this could be
realized in materials. To our opinion this is just to complete the
discussion, but without striking relevance for real systems.
As expected an increase of thermopower and figure of merit is
obtained. The behavior of $ZT$ will be discussed in detail in
fig.~\ref{fig:ZT-max-GG}. 
  \subsection{Simple estimate for S using {\sc Mott}'s Formula}
A very rough estimate for the {\bf Maximum Thermopower} can be
obtained using 
{\bf\sc Mott}'s formula: 
\begin{equation}
S \approx - \left. \frac{k_B}{|e|} 3 k_B T \frac{W '}{W}
\right|_{E=\mu}
\quad .
\end{equation}
Assuming a smooth behavior of $W(E)$
on a scale of $k_BT$ restricts the
logarithmic derivative $|W'/W|$ 
to about $1/k_BT$.
This limits the thermopower $|S|$ to 
{$\boldmath 3 \frac{k_B}{|e|}$}, 
which is about {\bf 250}~{\boldmath $\mu V/K$}.

So, by a assuming a smooth behavior of the transport spectral
function, the thermopower has a universal upper limit, which does not
depend on the character of the transport mechanism.

In the following we will calculate the thermopower in parallel to the
figure of merit using mainly $<\epsilon>$.
Assuming a broadening of the transport spectral function features by
the thermal energy scale $k_BT$ with a Lorentzian line shape, an upper
limit of 1.88 is obtained. Together with the thermopower quantum of
$k_B/|e|$ the upper limit is reduced to about 
{\bf 160}~{\boldmath $\mu V/K$}.

 \section{Results and Discussion}
     \subsection{Quantum-Dot System\label{res:1}}

%  Scheme 1

% S(E) Sys=1
\begin{figure}[hbt]
\centering
\includegraphics[width=0.9\columnwidth]{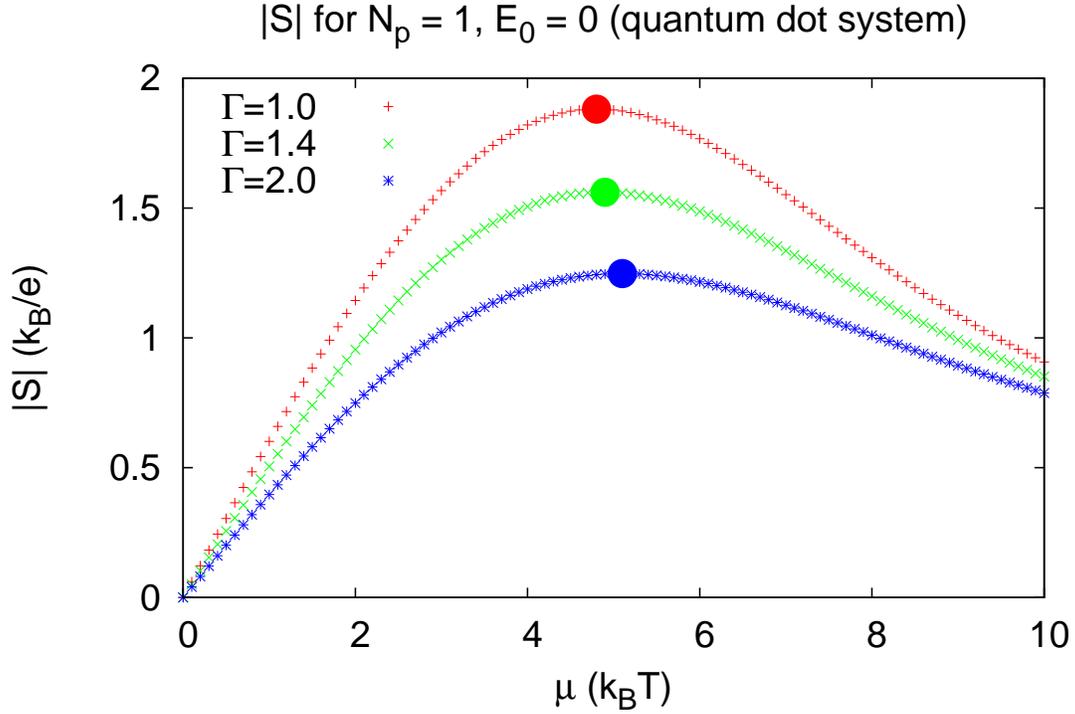}
\caption{\label{fig:S-E} 
Quantum dot system: Thermopower $|S|$ as function of chemical
Potential, 
The maximum $ZT$ for $\Gamma=1\times k_BT$ is 1.11.
}
\end{figure}

% ZT(E) Sys=1
\begin{figure}[hbt]
\centering
\includegraphics[width=0.9\columnwidth]{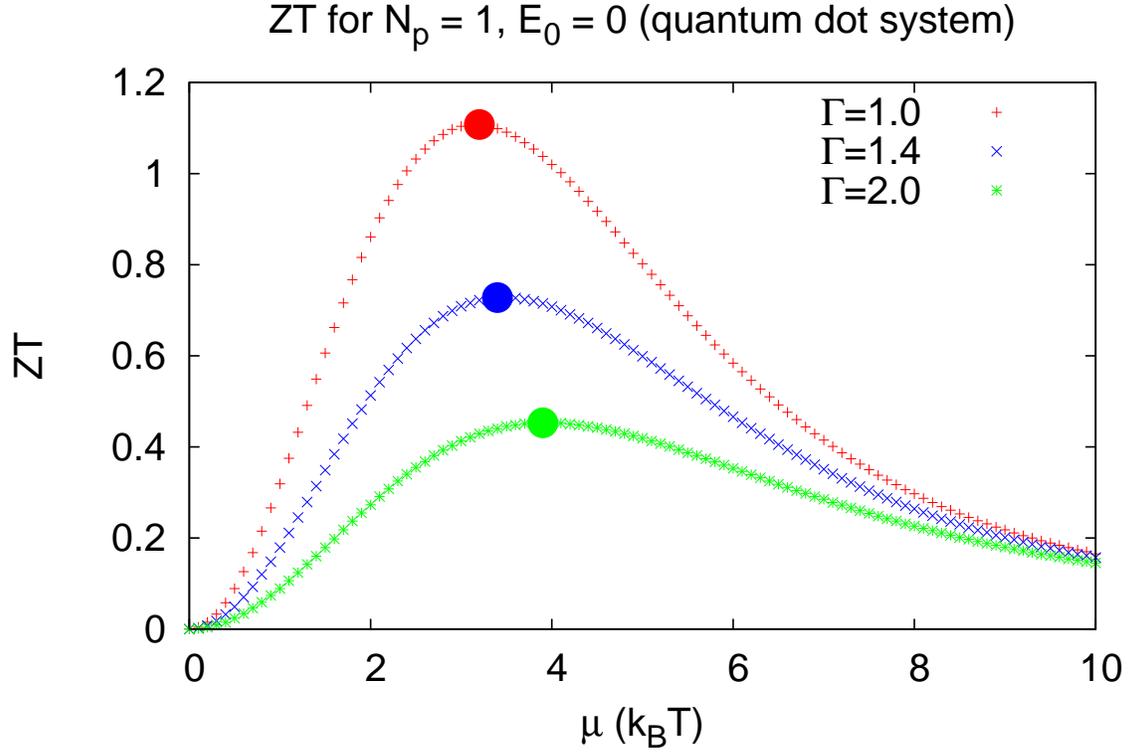}
\caption{\label{fig:ZT-E} 
Quantum dot system: The figure of merit $ZT$ as function of chemical
Potential.
The maximum $|S|$ for $\Gamma=1\times k_BT$ is 1.88.
}
\end{figure}

The main messages from these figures are:
The maximum of the thermopower |S| is obtained with 1.88 $k_B\|e|$ at a
chemical of about 4.8 $k_BT$. This is somewhat smaller than the
estimate obtained from the {\sc Mott} formula.
\\
The maximum figure of merit $ZT$ is obtained with about 1.11 at a
chemical potential of 3.2 $k_BT$. 
This universal value does not depend on the
character of the transport mechanism, nor the character of the current
as electron or hole.

\clearpage
  \subsection{2-level System\label{res:2}}

\begin{figure}[hbt]
\centering
\includegraphics[width=0.9\columnwidth]{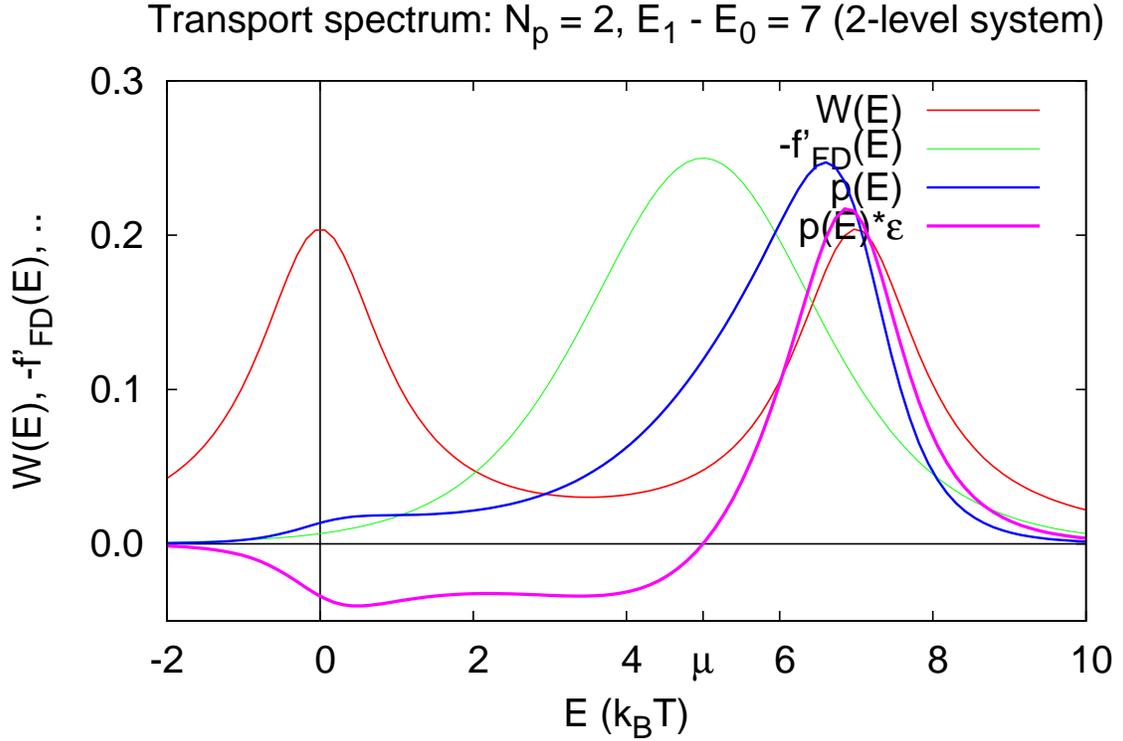}
\caption{\label{fig:s2}
2-level system system: 
Transport spectral function $W(E)$ for $E_1=7$, derivative of Fermi
occupation function $-f'_{FD}$ for $\mu=4.8$, the effective
distribution function $p(\epsilon)$, and the product
$p(\epsilon)\times\epsilon$ to visualize the contributions to
$<\epsilon>$. The energy $E$ and the relative energy $\epsilon=E-\mu$
are given in units of the thermal energy $k_BT$.
}
\end{figure}

\begin{figure}[hbt]
\centering
\includegraphics[width=0.9\columnwidth]{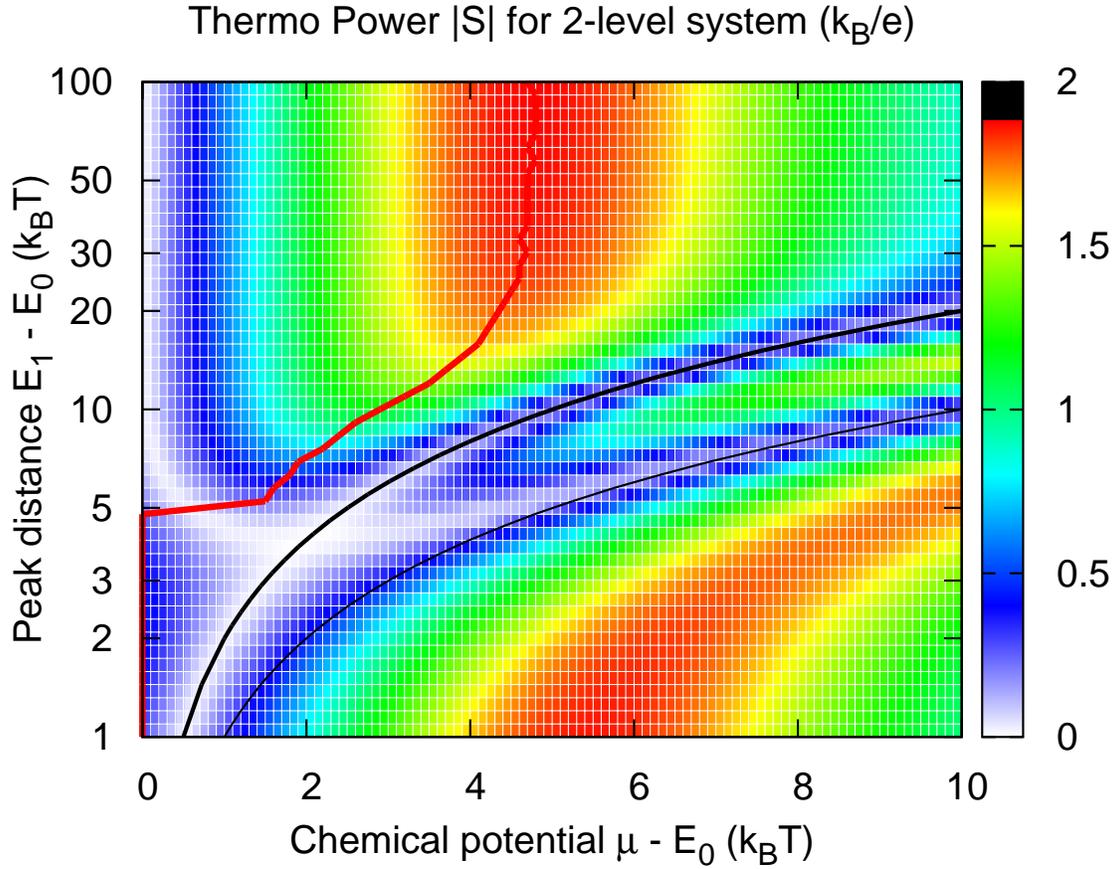}
\caption{\label{fig:S-E-E1} 
2-level system: The thermopower $|S|$ as function of peak distance
$E_1$ and chemical Potential $\mu$.
}
\end{figure}

\begin{figure}[hbt]
\centering
\includegraphics[width=0.9\columnwidth]{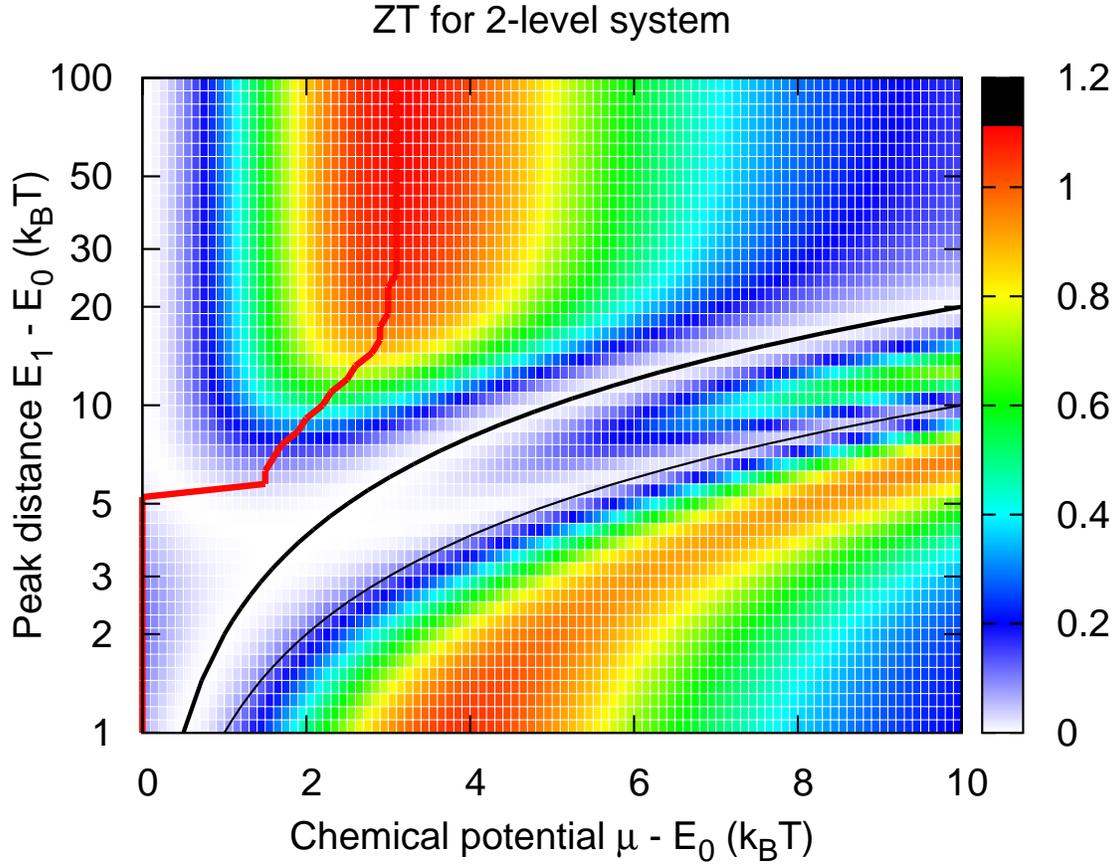}
\caption{\label{fig:ZT-E-E1} 
2-level system: The figure of merit $ZT$ as function of peak distance
$E_1$ and chemical Potential $\mu$.
}
\end{figure}

\begin{figure}[hbt]
\centering
\includegraphics[width=0.9\columnwidth]{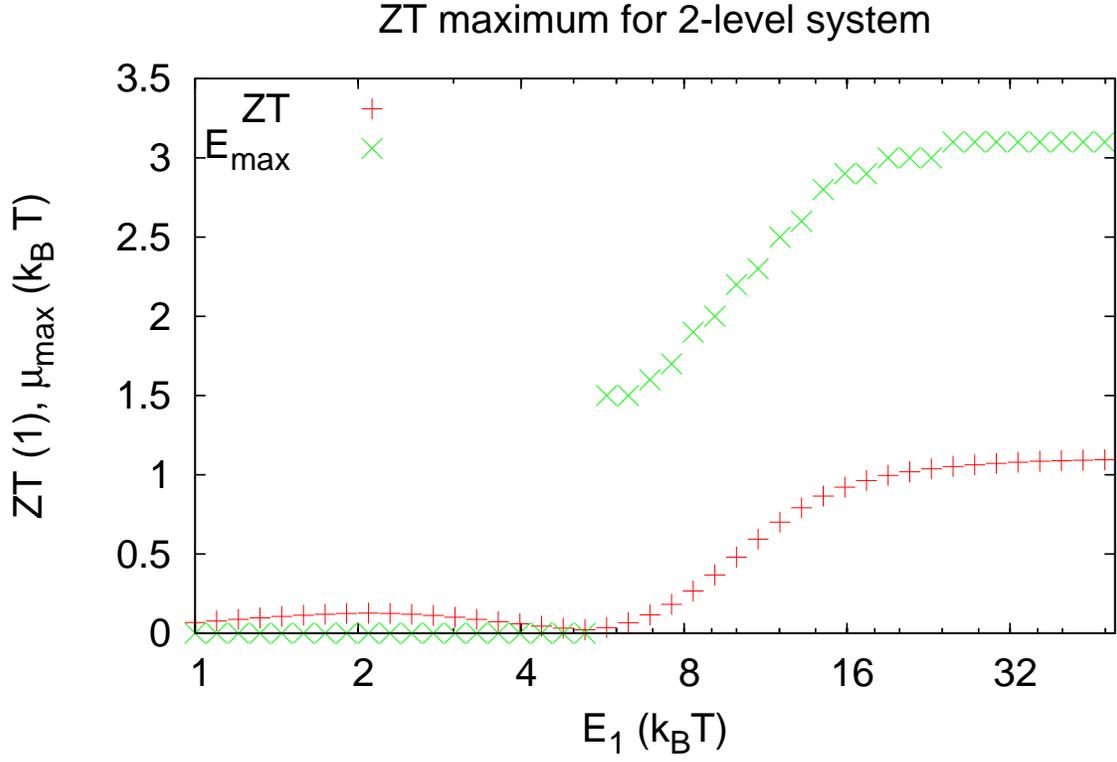}
\caption{\label{fig:ZT-max-E-E1} 
2-level system: Maximum figure of merit $ZT$ and position of chemical
potential $mu$ as function of peak distance $E_1$.
}
\end{figure}

Now we consider a 2-level system with symmetric peak weights, see
fig.~\ref{fig:s2}.
The additional lines in figs. \ref{fig:S-E-E1} and \ref{fig:ZT-E-E1}
show:
\\
- lower black solid line: position of the 2$^{nd}$ Lorentzian peak at
energy $E_1$
\\
- upper black solid line: middle of the "band gap" at $E_1/2$:
all results depending on energy $E$ are 
symmetric with respect to this line $E_1/2$,
\\
- red line: position of $ZT$ maximum depending on the 'band gap'
$E_1$, 
for this, only values of $\mu$ inside the 'band gap' were considered,
so between $0$ and$E_1/2$. The behavior of the maximum $ZT$ value and the
corresponding chemical potential is analyzed in more detail in
fig. \ref{fig:ZT-max-E-E1}. 

% S(E) Sys=2

% ZT(E) Sys=2

The behavior of $ZT$ as function of the peak position (the effective
band gap) shows 2 regions:
For small peak separations, $ZT$ is very small and the optimum is
obtained for chemical potential positions very close to one of the
peak centers- here shown for the left peak.
For larger separations, $ZT$ tends to it's maximum value and the
optimum $\mu$ is about 3.2 in units of $k_BT$.
This transition occurs for peak separations between 8 and 16 $k_BT$.

% ZT_max(E) Sys=2

\clearpage
  \subsection{Asymmetric 2-level system\label{res:3}}

Now, the asymmetric 2-level system will be analyzed:
\\
- the maximum values for 
$|S|$ and $ZT$ for systems
with different peak heights 
do not exceed the values found for 
the quantum dot and 
the symmetric 2-level system,
\\
- the largest values are obtained 
for small distances - 
close to the single peak case /1/,
and for large peak separations - 
similar to case /2/,
the case for large separations $E_1$ is not shown 
in the figures.

\begin{figure}[hbt]
\centering
\includegraphics[width=0.9\columnwidth]{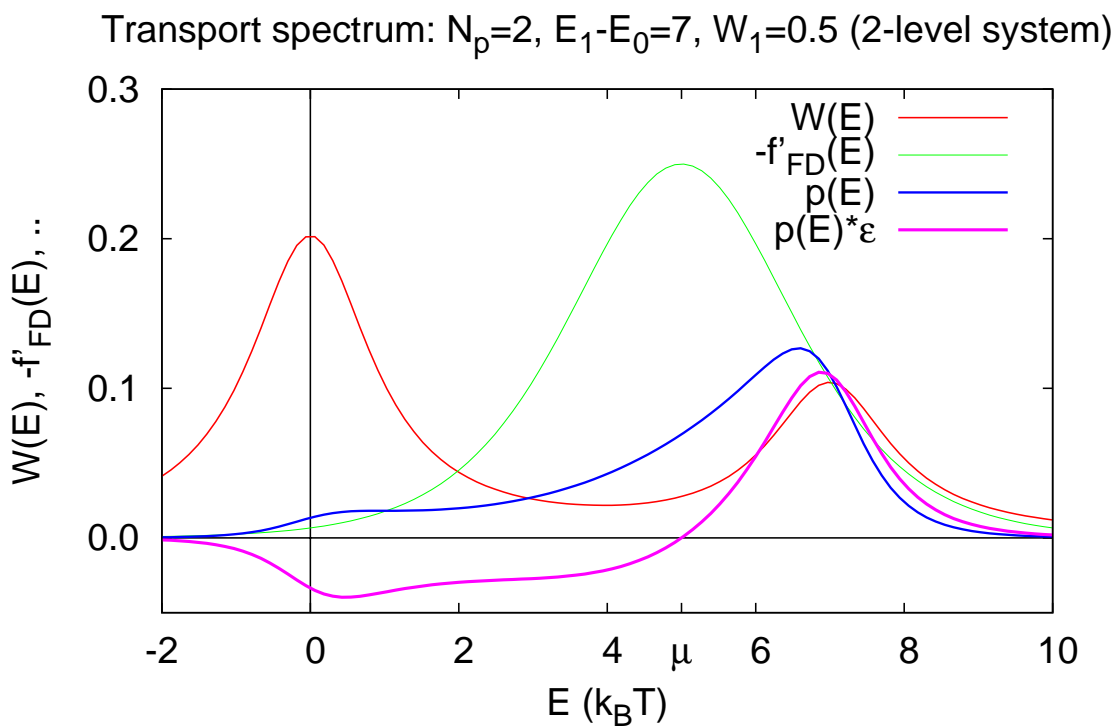}
\caption{\label{fig:s3} 
Asymmetric 2-level system: 
Transport spectral function $W(E)$ for $E_1=7$, $W_0=1$, and $W_1=.5$, 
derivative of Fermi
occupation function $-f'_{FD}$ for $\mu=4.8$, the effective
distribution function $p(\epsilon)$ and the product
$p(\epsilon)\times\epsilon$ to visualize the contributions to
$<\epsilon>$. The energy $E$ and the relative energy $\epsilon=E-\mu$
are given in units of the thermal energy $k_BT$.
}
\end{figure}

\begin{figure}[hbt]
\centering
\includegraphics[width=0.9\columnwidth]{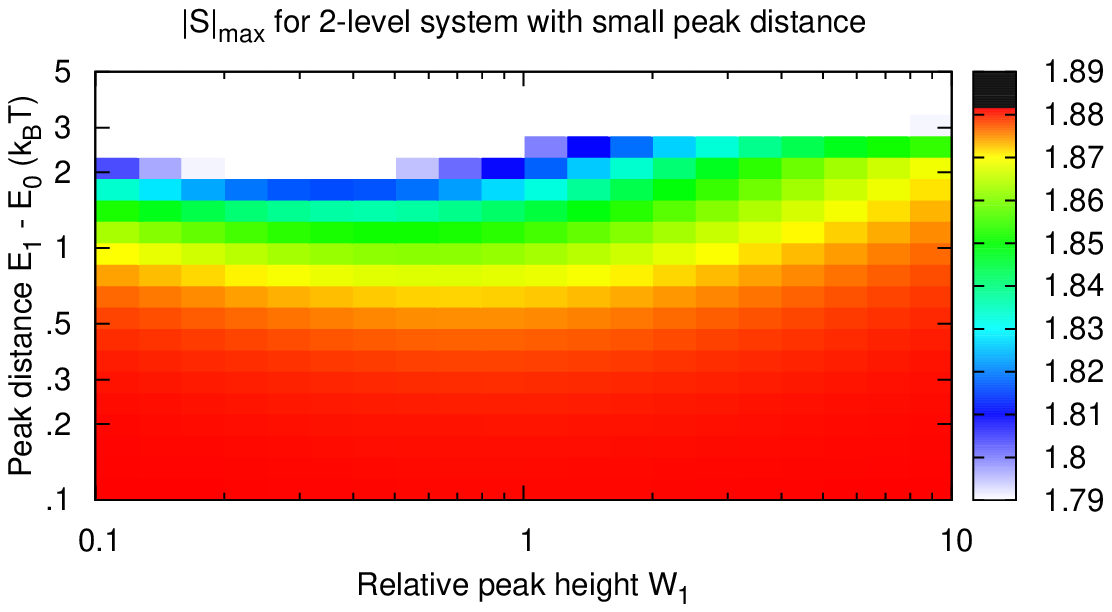}
\caption{\label{fig:S-max-W1-E1} 
Asymmetric 2-level system: 
$S$ depending on $E_1$, $W_0=1$, and $W_1=.5$, 
derivative of Fermi
occupation function $-f'_{FD}$ for $\mu=4.8$, the effective
distribution function $p(\epsilon)$ and the product
$p(\epsilon)\times\epsilon$ to visualize the contributions to
$<\epsilon>$. The energy $E$ and the relative energy $\epsilon=E-\mu$
are given in units of the thermal energy $k_BT$.
}
\end{figure}

\begin{figure}[hbt]
\centering
\includegraphics[width=0.9\columnwidth]{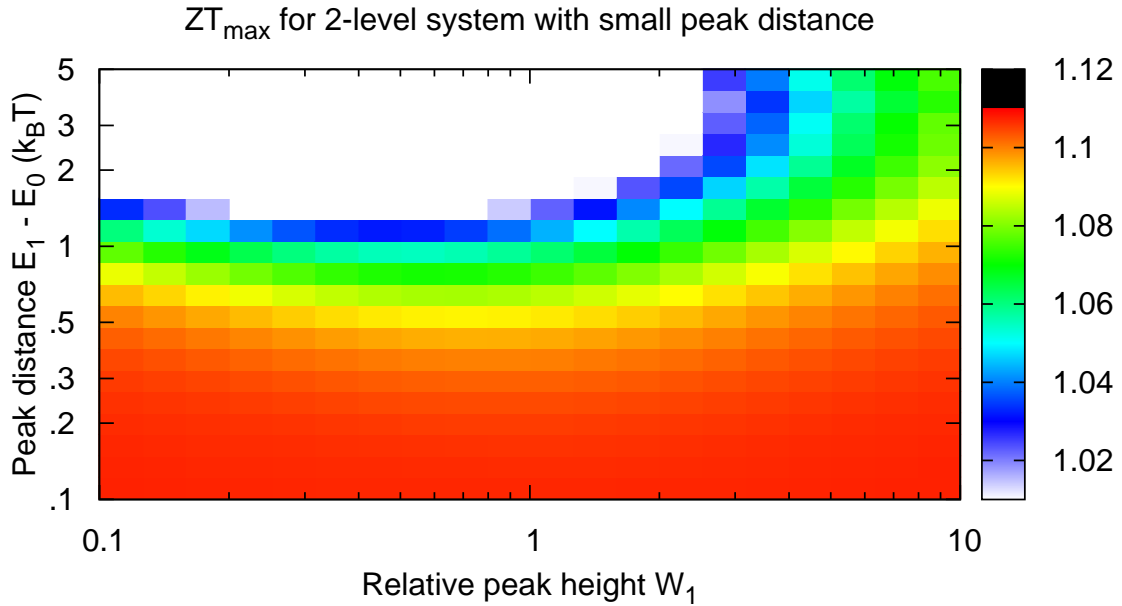}
\caption{\label{fig:ZT-max-W1-E1} 
Asymmetric 2-level system: 
$ZT$ as function of $E_1$, and relative peak width $W_1$.
}
\end{figure}
\clearpage
  \subsection{Scaled Electron-Phonon Coupling\label{res:4}}
Here the case of smaller peak width than the thermal energy is
considered.
As expected a strong increase of the maximum $ZT$ is obtained.
\begin{figure}[hbt]
\centering
\includegraphics[width=0.9\columnwidth]{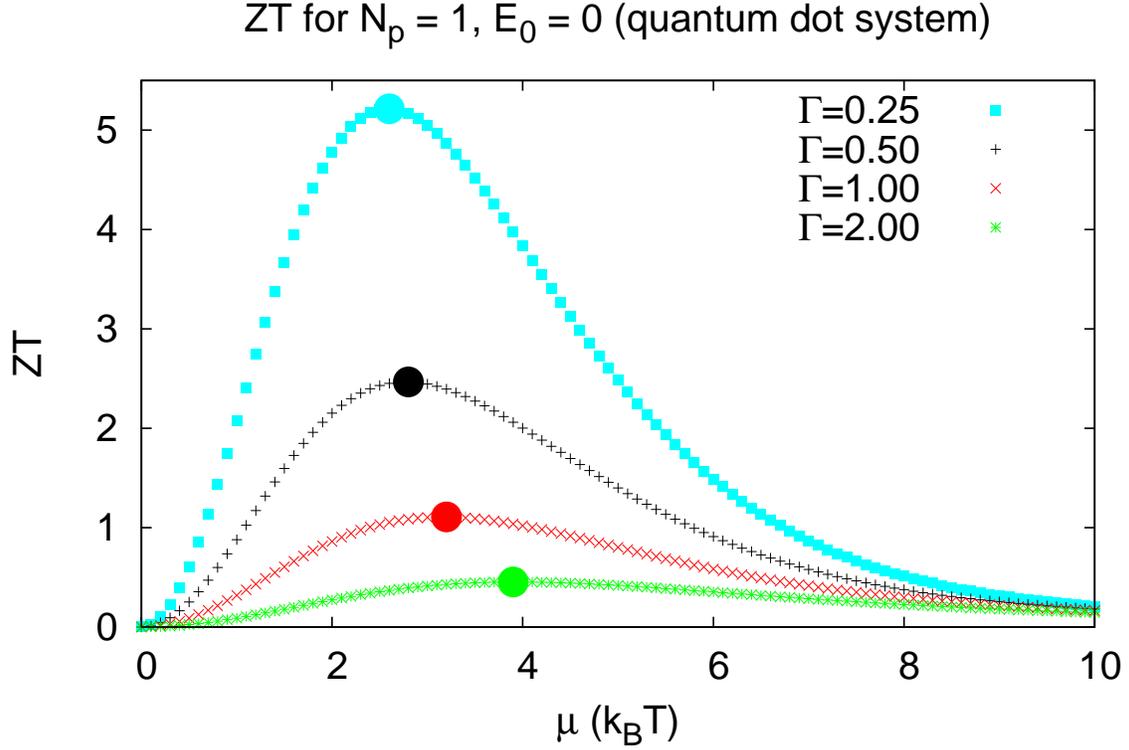}
\caption{\label{fig:ZT-GG}
Quantum dot system: 
$ZT$ as function of chemical potential $\mu$ for different effective
coupling strengths $\Gamma$.
The maximum of $ZT$ is marked by dots. The maximum values are
summarized in Fig.\ref{fig:ZT-max-GG}.
}
\end{figure}

Assuming a weaker broadening than $\Gamma=k_BT$, 
$ZT$ increases stronger than inverse proportional, 
A $\Gamma^{-5/4}$ behavior is roughly obtained.

\clearpage

 \section{Conclusions}
  % Conclusions
%
To conclude these considerations we state that for a large ZT
the following three conditions have to be met by the transport
mechanism in the thermoelectric material:
\\
1. The band gap has to be quite large, at least larger than 8
times $k_BT$.
\\
2.To suppress the lattice thermal conductivity a phonon glass like
behavior should be present.
\\
3. The electron-phonon interaction should be smallest as possible.
To our understanding, a minimum peak width in the transport spectral
function of $k_BT$ can be realized.
\\
Independent on the transport mechanisms the following universal limits
are obtained:
\\
The absolute value of the thermopower $|S|$ is limited by about 160 $\mu
V/K$.
\\
The figure of merit $ZT$ is limited by about 1.11.

 \section*{Acknowledgments}
  We thank S. Gemming for fruitful discussions.
This work has been partially financed by the Initiative and Networking
Fund of the German Helmholtz Association via the Helmholtz International
Research School NanoNet \mbox{(VH-KO-606)} and the Helmholtz
Exzellenznetzwerk cfaed \mbox{(ExNet-0026)}. 
We gratefully acknowledge partial funding by the DFG via Focus
Research Program SSP 1386.

 \clearpage
 \section*{References}
 \bibliography{lit-ZT}

\begin{thebibliography}{8}
\expandafter\ifx\csname natexlab\endcsname\relax\def\natexlab#1{#1}\fi
\expandafter\ifx\csname bibnamefont\endcsname\relax
  \def\bibnamefont#1{#1}\fi
\expandafter\ifx\csname bibfnamefont\endcsname\relax
  \def\bibfnamefont#1{#1}\fi
\expandafter\ifx\csname citenamefont\endcsname\relax
  \def\citenamefont#1{#1}\fi
\expandafter\ifx\csname url\endcsname\relax
  \def\url#1{\texttt{#1}}\fi
\expandafter\ifx\csname urlprefix\endcsname\relax\def\urlprefix{URL }\fi
\providecommand{\bibinfo}[2]{#2}
\providecommand{\eprint}[2][]{\url{#2}}

\bibitem[{\citenamefont{Eltschka et~al.}(2016)\citenamefont{Eltschka,
  Thierschmann, Buhmann, and Siewert}}]{eltschka16}
\bibinfo{author}{\bibfnamefont{C.}~\bibnamefont{Eltschka}},
  \bibinfo{author}{\bibfnamefont{H.}~\bibnamefont{Thierschmann}},
  \bibinfo{author}{\bibfnamefont{H.}~\bibnamefont{Buhmann}}, \bibnamefont{and}
  \bibinfo{author}{\bibfnamefont{J.}~\bibnamefont{Siewert}},
  \bibinfo{journal}{Phys. Stat. Sol. A} \textbf{\bibinfo{volume}{13}},
  \bibinfo{pages}{626} (\bibinfo{year}{2016}).

\bibitem[{\citenamefont{Mahan and Sofo}(1996)}]{mahan96}
\bibinfo{author}{\bibfnamefont{G.}~\bibnamefont{Mahan}} \bibnamefont{and}
  \bibinfo{author}{\bibfnamefont{J.}~\bibnamefont{Sofo}},
  \bibinfo{journal}{Proc. Nat. Acad. Sci.} \textbf{\bibinfo{volume}{93}},
  \bibinfo{pages}{7436} (\bibinfo{year}{1996}).

\bibitem[{\citenamefont{Matveev}(2000)}]{matveev00}
\bibinfo{author}{\bibfnamefont{K.}~\bibnamefont{Matveev}}, in
  \emph{\bibinfo{booktitle}{Statistical and Dynamical Aspects of Mesoscopic
  Systems, Springer Lecture Notes in Physics, Vol. 547}}, edited by
  \bibinfo{editor}{\bibfnamefont{D.}~\bibnamefont{Reguera}},
  \bibinfo{editor}{\bibfnamefont{G.}~\bibnamefont{Platero}},
  \bibinfo{editor}{\bibfnamefont{L.~L.} \bibnamefont{Bonilla}},
  \bibnamefont{and} \bibinfo{editor}{\bibfnamefont{J.~M.} \bibnamefont{Rubi}}
  (\bibinfo{publisher}{Springer}, \bibinfo{address}{Berlin},
  \bibinfo{year}{2000}).

\bibitem[{\citenamefont{Vining}(2009)}]{vining09}
\bibinfo{author}{\bibfnamefont{C.~B.} \bibnamefont{Vining}},
  \bibinfo{journal}{Nature Mat.} \textbf{\bibinfo{volume}{8}},
  \bibinfo{pages}{83} (\bibinfo{year}{2009}).

\bibitem[{\citenamefont{Harman et~al.}(2005)\citenamefont{Harman, Walsh,
  Laforge, and Turner}}]{harman05}
\bibinfo{author}{\bibfnamefont{T.~C.} \bibnamefont{Harman}},
  \bibinfo{author}{\bibfnamefont{M.~P.} \bibnamefont{Walsh}},
  \bibinfo{author}{\bibfnamefont{B.~E.} \bibnamefont{Laforge}},
  \bibnamefont{and} \bibinfo{author}{\bibfnamefont{G.~W.~J.}
  \bibnamefont{Turner}}, \bibinfo{journal}{Electr. Mat.}
  \textbf{\bibinfo{volume}{34}}, \bibinfo{pages}{L19} (\bibinfo{year}{2005}).

\bibitem[{\citenamefont{Venkatasubramanian
  et~al.}(2001)\citenamefont{Venkatasubramanian, Siilova, Colpitts, and
  O'Quinn}}]{rama01}
\bibinfo{author}{\bibfnamefont{R.}~\bibnamefont{Venkatasubramanian}},
  \bibinfo{author}{\bibfnamefont{E.}~\bibnamefont{Siilova}},
  \bibinfo{author}{\bibfnamefont{T.}~\bibnamefont{Colpitts}}, \bibnamefont{and}
  \bibinfo{author}{\bibfnamefont{B.}~\bibnamefont{O'Quinn}},
  \bibinfo{journal}{Nature} \textbf{\bibinfo{volume}{413}},
  \bibinfo{pages}{597} (\bibinfo{year}{2001}).

\bibitem[{\citenamefont{Hsu et~al.}(2004)}]{hsu04}
\bibinfo{author}{\bibfnamefont{K.~F.} \bibnamefont{Hsu}} \bibnamefont{et~al.},
  \bibinfo{journal}{Science} \textbf{\bibinfo{volume}{303}},
  \bibinfo{pages}{818} (\bibinfo{year}{2004}).

\bibitem[{\citenamefont{Hinsche et~al.}(2016)\citenamefont{Hinsche, Rittweger,
  H{\"o}lzer, Zahn, Ernst, and Mertig}}]{hinsche16}
\bibinfo{author}{\bibfnamefont{N.~F.} \bibnamefont{Hinsche}},
  \bibinfo{author}{\bibfnamefont{F.}~\bibnamefont{Rittweger}},
  \bibinfo{author}{\bibfnamefont{M.}~\bibnamefont{H{\"o}lzer}},
  \bibinfo{author}{\bibfnamefont{P.}~\bibnamefont{Zahn}},
  \bibinfo{author}{\bibfnamefont{A.}~\bibnamefont{Ernst}}, \bibnamefont{and}
  \bibinfo{author}{\bibfnamefont{I.}~\bibnamefont{Mertig}},
  \bibinfo{journal}{Phys. Stat. Sol. A} \textbf{\bibinfo{volume}{13}},
  \bibinfo{pages}{672} (\bibinfo{year}{2016}).

\end{thebibliography}
\end{document}